\documentclass[bibyear]{aa}

\usepackage{graphicx}
\usepackage{txfonts}
\usepackage{natbib}
\bibpunct{(}{)}{;}{a}{}{,} 
\usepackage{booktabs}
\usepackage{multicol}
\usepackage{graphicx}
\usepackage{fancyhdr}
\usepackage{amsmath}	
\usepackage{amssymb}	
\usepackage[inter-unit-product=\cdot]{siunitx}
\usepackage{enumerate}
\usepackage{multirow}
\usepackage{mathtools}
\usepackage{longtable}
\usepackage{tabularx}
\usepackage{xcolor} 
\usepackage{ulem} 
\usepackage{comment}
\usepackage{cuted}
\usepackage{soul}

\usepackage{hyperref}
\hypersetup{colorlinks=true,urlcolor=black, citecolor=blue, linkcolor=black}

\makeatletter
\renewcommand*\aa@pageof{, page \thepage{} of \pageref*{LastPage}}
\makeatother

\newcommand{\orcidicon}[1]{\href{https://orcid.org/#1}{\includegraphics[width=11pt]{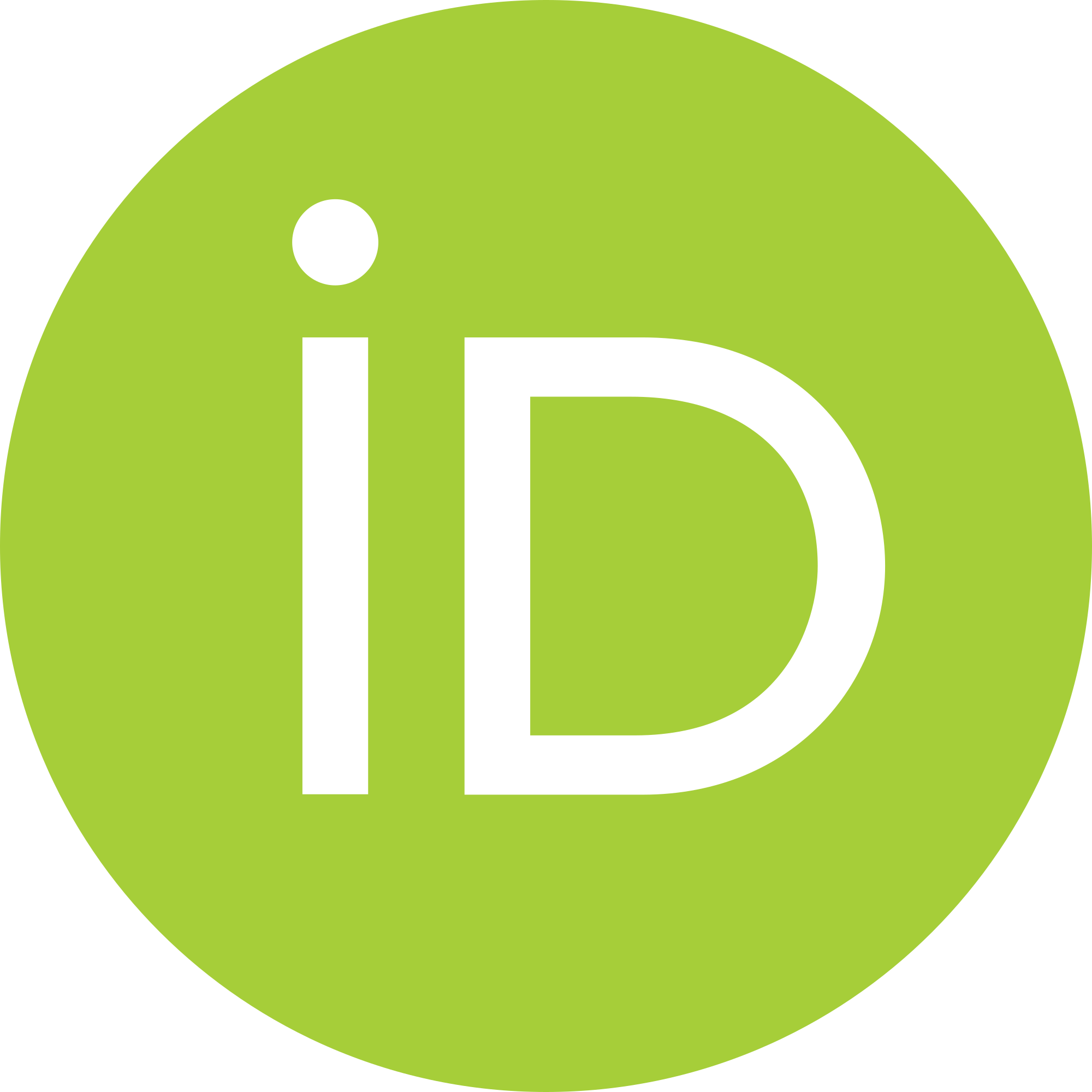}}}
\newcommand{\orcid}[1]{\href{https://orcid.org/#1}{\protect\orcidicon{#1}}}

\newcommand{\msun}{{\rm M}_\odot}

\definecolor{steelblue}{rgb}{0.274 0.510 0.706}

\defcitealias{Sabhahit2022}{S22}
\defcitealias{Sabhahit2023}{S23}
\defcitealias{Chen2015}{C15}

\begin{document}

   \title{Optically thick winds of very massive stars \\ suppress intermediate-mass black hole formation}
    \titlerunning{Optically thick winds suppress IMBH formation}
   
   \author{
    Stefano Torniamenti\inst{1,2},  
    \orcid{0000-0002-9499-1022} \thanks{\href{mailto:stefano.torniamenti@unipd.it}{sttorniamenti@mpia.de}}, 
    Michela Mapelli\inst{2,3,4,5}\orcid{0000-0001-8799-2548}\thanks{\href{mailto:mapelli@uni-heidelberg.de}{mapelli@uni-heidelberg.de}},
    Lumen Boco\inst{2}
    \orcid{0000-0003-3127-922X} \thanks{\href{mailto:lumen.boco@uni-heidelberg.de}{lumen.boco@uni-heidelberg.de}}, \\
    Filippo Simonato \inst{4,6,7},
    Giuliano Iorio\inst{8}\orcid{0000-0003-0293-503X}, 
    Erika Korb\inst{4,2,5}\orcid{0009-0007-5949-9757}
    }
    \authorrunning{S. Torniamenti et al.}
    \institute{
    $^{1}$Max-Planck-Institut f{\"u}r Astronomie, K{\"o}nigstuhl 17, 69117, Heidelberg, Germany\\
    $^{2}$Universit\"at Heidelberg, Zentrum f\"ur Astronomie (ZAH), Institut f\"ur Theoretische Astrophysik, Albert Ueberle Str. 2, 69120, Heidelberg, Germany\\
    $^3$Universit\"at Heidelberg, Interdiszipli\"ares Zentrum f\"ur Wissenschaftliches Rechnen, D-69120 Heidelberg, Germany\\
    $^4$Dipartimento di Fisica e Astronomia Galileo Galilei, Università di Padova, Vicolo dell’Osservatorio 3, I–35122 Padova, Italy\\
    $^5$INFN, Sezione di Padova, Via Marzolo 8, I--35131 Padova, Italy\\
    $^6$Gran Sasso Science Institute (GSSI), Viale Francesco Crispi 7, 67100 L’Aquila, Italy\\
    $^7$INFN, Laboratori Nazionali del Gran Sasso, 67100 Assergi, Italy\\
    $^8$Departament de Física Quàntica i Astrofísica, Institut de Ciències del Cosmos, Universitat de Barcelona, Martí i Franquès 1, E-08028 Barcelona, Spain\\
    }
   \date{Received XXXX; accepted YYYY}

\abstract{
Intermediate-mass black holes (IMBHs) are the link between stellar-mass and supermassive black holes.  
Gravitational waves have 
started unveiling a population of IMBHs in the $\sim 100-300 \, \msun$ range.  
Here, we investigate the formation of IMBHs from {non-rotating} very massive stars (VMSs, $>100\,{} \msun$). We calculate new VMS models that account for the transition from optically thin to optically thick winds, and study how this enhanced mass loss affects  IMBH formation and the black hole mass function at intermediate and high metallicity ($Z=10^{-4}-0.02$). 
We show that optically thick winds suppress the formation of IMBHs from direct VMS collapse at metallicities $Z>0.001$, one order of magnitude lower than  predicted by previous models. 
Our models indicate that the stellar progenitors of GW231123 must have had a metallicity $Z<0.002$, if the primary black hole formed via direct VMS collapse. 
}
  

   \maketitle
%

\section{Introduction}

Intermediate-mass black holes (IMBHs), with masses $M_{\mathrm{BH}}\sim{10^2-10^5}$~M$_\odot$ are the link between stellar-mass and supermassive black holes at the center of galaxies \citep{greene2020}. In the last few years, gravitational waves have provided the first compelling evidence of their existence. The merger remnant of GW190521 represents the first IMBH ever detected via gravitational waves, with a mass of $142^{+28}_{-16} \, \mathrm{M_{\odot}}$ \citep{abbottGW190521,abbottGW190521astro}. More recently, parameter estimation for the event GW231123 \citep{gw231123} reveals a black hole (BH) remnant with $m \sim 240 \, \msun$. Both BBH components of GW231123 have masses within or above the pair-instability mass gap \citep{Heger2002,Woosley2007}, with peculiarly high spins ($\chi \gtrsim 0.7$) that hint at the dynamical origin of this merger \citep{gerosa2017,fishbach2017,antonini2019,gerosa2021,mapelli2021,antonini2025, kiroglu2025b,kiroglu2025}. The current fourth gravitational-wave transient catalog \citep{GWTC4pop,GWTC4} already contains a number of other IMBH candidates, which is expected to increase with the next observing runs.

At the same time, searches for IMBHs in the Milky Way globular clusters have mostly led to inconclusive results \citep{mezcua2017}, with only a few debated candidate BHs \citep{millerjones2012,tremou2018,baumgardt2019}. One notable exception is represented by $\omega$ Centauri, which hosts an IMBH with mass $\ge{} 8\times 10^3 \, \msun$, as revealed by proper motions based on  over 20 years of HST data \citep{haberle2024}. 

Possible IMBH formation pathways include the direct collapse of very massive stars (VMS), with masses above 100~M$_\odot$ \citep{Belkus2007,Crowther2010, Bestenlenher2014, vink2015,costa2025,shepherd2025}, stellar collisions in dense young star clusters \citep{Portegies2002,freitag2006a,freitag2006b,vanbeveren2009,mapelli2016,dicarlo2021,torniamenti2022}, and hierarchical BH mergers in dense and massive star clusters \citep{antonini2019,antonini2023,mapelli2021, rizzuto2021,arcasedda2023,torniamenti2024}. 

VMSs have been observed in star forming regions and young massive clusters \citep{vink2015}, like the Arches cluster near the Galactic Center \citep{martins2008}, NGC 3603 \citep{Crowther2010}, and R136 in the Large Magellanic Cloud (LMC, \citealp{Crowther2010,Bestenlenher2014,bestenlehner2020}). 
The estimated initial masses of these objects are up to $\sim 300 \, \msun$ \citep{Crowther2010}. Also, most of them are observed as single stars \citep{Crowther2010}, which may hint that their formation is triggered or enhanced by stellar collisions during the earliest stellar phases \citep{portegieszwart1999,mapelli2016}. The possibility that VMSs form IMBHs, however, strongly depends on the amount of mass they lose during their life \citep{vink2018}.

At solar metallicity, VMSs can lose a large fraction of their initial mass already during the main sequence, and leave relatively low-mass compact objects \citep[$<30$~M$_\odot$, ][]{belczynski2010,mapelli2013,romagnolo2024}. 
At low metallicity, these stars could preserve enough mass to enter the pair-instability regime \citep{Spera2017,renzo2020}. Namely, stars developing He cores in the 64--135~M$_\odot$ range at the end of carbon burning  eventually explode as (pair-instability) supernovae leaving no compact remnant \citep{Heger2002,Woosley2007, yungelson2008}, whereas stars that develop even more massive He cores efficiently undergo photodisintegration and possibly collapse into IMBHs \citep{Spera2017}. 
Also, the retention of the envelope can trigger a dredge-up phase during the core helium-burning phase, remove mass from the stellar core and, in turn, prevent pair-production episodes \citep{Costa2021}. 

Most current models predict that VMSs can form IMBHs up to metallicities $Z \lesssim 0.014$ (see, e.g., \citealp{costa2025}), where stellar winds become strong enough to prevent the formation of stellar cores with $> 100 \, \msun$. 
Recently, \citet[][hereafter \citetalias{Sabhahit2022, Sabhahit2023}]{Sabhahit2022,Sabhahit2023} introduced a new model for stellar winds in VMSs based on the concept of transition mass loss \citep{Vink2011,Vink2012} to account for the transition from optically thin winds of O-type stars to the optically thick wind regime. This predicts {enhanced mass-loss rates ($\dot{M}\gtrsim{}10^{-4}$~M$_\odot \, \mathrm{yr^{-1}}$) already during the main sequence,} when the star is radiation-pressure dominated -- i.e., when its luminosity is very close to or even above the Eddington value. 
In this case, the wind mass-loss rate is strong enough to reduce the core mass (\citetalias{Sabhahit2023}). This improved wind implementation has proved to naturally account for the narrow range of observed VMS temperatures in the Galaxy and the LMC (\citetalias{Sabhahit2022}). Also, it heavily suppresses the occurrence of pair instability due to lower core masses. In a recent work, we estimate that this new wind formalism can naturally yield rates of pair instability supernovae consistent with observations \citep{simonato2025}. Another consequence we may naturally expect is the suppression of IMBH production from VMSs in the regime where optically thick winds become effective.
 
Here, we investigate the formation of IMBHs from VMSs. We account for optically thick winds to assess how different metallicities and mass-loss rates affect the formation of these objects. To this purpose, we have run an extensive set of VMS 
models with \textsc{mesa} \citep{Paxton2011,Paxton2018} to explore the impact of stellar winds on the mass of the VMS cores and the resulting BH masses. We find that IMBH formation is strongly suppressed by our optically thick wind model, even at the metallicity of the LMC.  

This paper is organized as follows. In Sec. \ref{sec:methods} we introduce our models for stars and stellar winds. In Sec. \ref{sec:results}, we show the BH mass distributions from single and binary stars in presence of optically thick winds. {In Sect.~\ref{sec:discussion} we discuss the possible impact of rotation and core overshooting.} Finally we summarize our results and conclusion in Sec. \ref{sec:summary}.

\section{Methods} \label{sec:methods}
We model the evolution of VMSs with the hydro-static stellar evolution code \textsc{mesa} (version r12115; \citealp{Paxton2011, Paxton2013, Paxton2015, Paxton2018, Paxton2019,jermy2023}) adopting the radiation-driven wind models by \citetalias{Sabhahit2023}. First, we evolve VMSs until the last phases before core collapse. Then, we use the population-synthesis code \textsc{sevn} \citep{Spera2019,Mapelli2020,Iorio2023} to explore the BH mass distribution of single and binary VMSs with different wind models and to study a population of massive binary stars. In the following, we introduce our initial conditions and model parameters.

\begin{figure*}
    \centering
    \includegraphics[width=\textwidth]{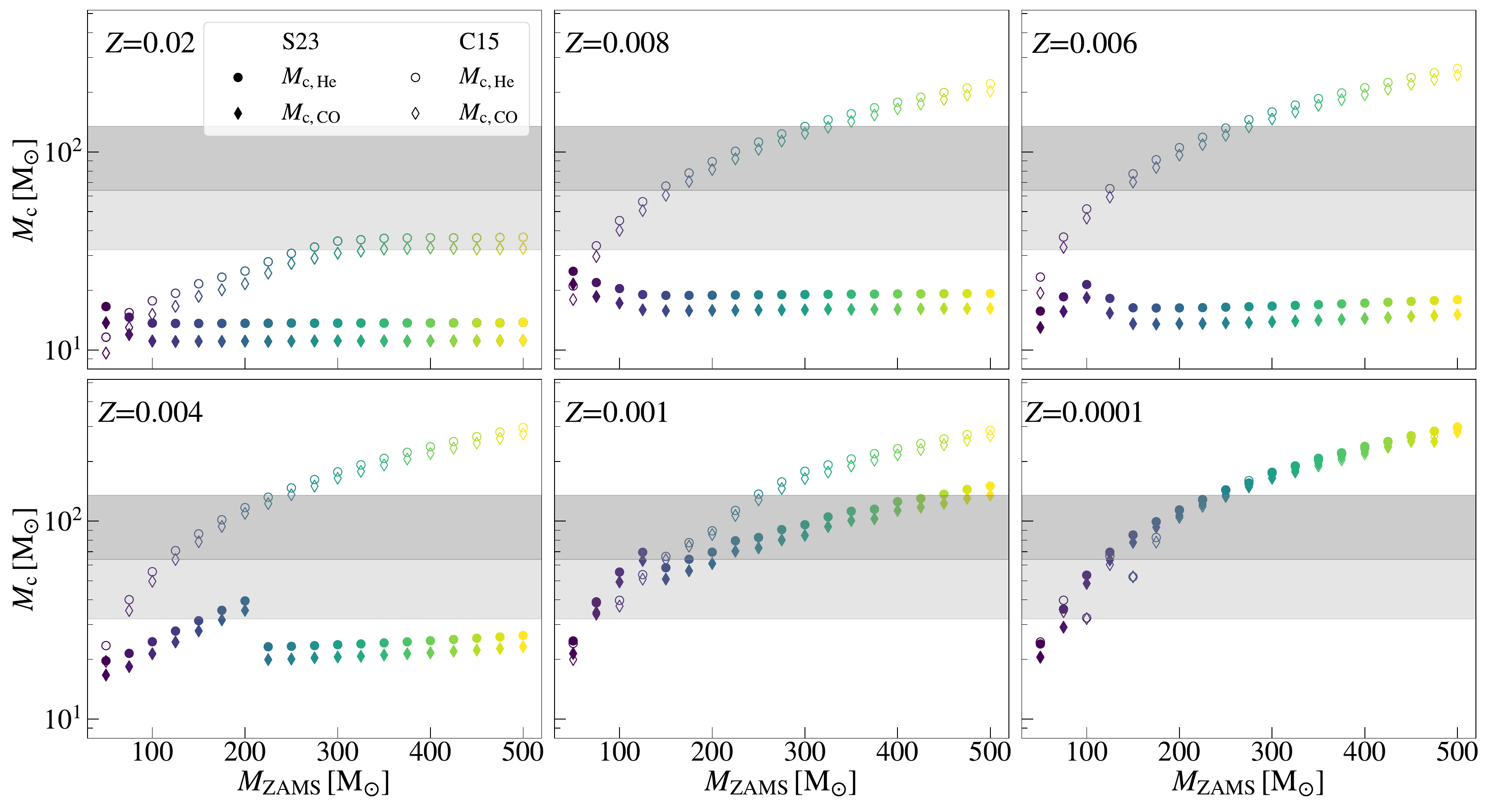}
    \caption{Masses of the He (circles) and CO cores (diamonds) for the wind model from \protect\citetalias{Sabhahit2023} (full markers) and \protect\citetalias{Chen2015} (void markers). The dark (light) gray shaded areas represent the regime for (pulsational) pair-instability supernovae.}
    \label{fig:pisn_new}
\end{figure*}

\subsection{VMS models with \textsc{mesa}}\label{sec:mesa}
We adopt the same initial set-up as \cite{simonato2025}, based on the wind mass-loss scheme by \citetalias{Sabhahit2022} and \citetalias{Sabhahit2023}\footnote{The \textsc{mesa} inlists and run\_star\_extras files by \cite{Sabhahit2023} are publicly available at \href{{https://github.com/Apophis-1/}}{this link}.}.  
This wind model is based on the concept of  transition mass loss introduced by \cite{Vink2012}: the transition from optically thin to optically thick winds takes place when the Eddington ratio of the star ($\Gamma_{\mathrm{Edd}}$) exceeds the ratio at the transition point  $\Gamma_{\mathrm{Edd, tr}}$. At each iteration, we calculate the mass-loss rate $\dot{M}_{\mathrm{Vink}}$, the escape velocity $v_{\mathrm{esc}}$ and the terminal speed $v_{\infty}$ from the stellar luminosity as described in \citetalias{Sabhahit2023}. 
We use these quantities to calculate the wind efficiency parameter, which determines when the switch occurs:
\begin{equation}
    \eta_{\mathrm{Vink}}(L) \equiv \frac{\dot{M}_{\mathrm{Vink}} \,v_{\infty}}{L / c}\, = \frac{\eta}{\tau_{F,s}}\left(\frac{v_{\infty}}{v_{\rm{esc}}}\right) = 0.75 \left(1 + \frac{v_{\rm{esc}}^2}{v_{\infty}^2}(L)\right)^{-1}\,,
\end{equation}
where $\tau_{F,s}$ is the  the flux-weighted mean optical depth at the sonic point.
In the optically thick regime, the wind mass-loss follows \cite{Vink2011}:
\begin{equation}\label{eq:mloss}
    \dot{M} = \dot{M}_{\mathrm{tr}}\,\left(\frac{L}{L_{\mathrm{tr}}}\right)^{4.77}\,\left(\frac{M}{M_{\mathrm{tr}}}\right)^{-3.99}\,,
\end{equation}
where $\dot{M}_{\mathrm{tr}}$, $L_{\mathrm{tr}}$ and $M_{\mathrm{tr}}$ are the mass-loss rate, the luminosity, and the mass at the transition point, respectively. 
During the main sequence, we use Eq.~\ref{eq:mloss} to describe the wind mass-loss in the optically thick regime. Otherwise, we refer to \cite{Vink2001}. In the core He-burning phase, we use Eq.~\ref{eq:mloss} for temperatures between $4\times10^3 \, \mathrm{K}$ and $10^5$ K, while for $T<4\times10^3 \, \mathrm{K}$, we use the formalism by \citet{deJager1988}. For $T>10^5$ K and central hydrogen mass fraction $X_{\text{C}}<0.01$, we adopt the Wolf-Rayet mass-loss by \citet{Sander2020}. 

We model convective mixing with the mixing length theory \citep{Cox1968}, adopting a constant mixing-length ratio $\alpha_{\text{MLT}} = 1.5$. We determine the convective boundaries by applying the Ledoux criterion \citep{Ledoux1947}. We use the semi-convective diffusion parameter $\alpha_{\text{SC}} = 1$.  We choose the exponential overshooting formalism by \cite{Herwig2000}, with efficiency parameter $f_{\text{OV}} = 0.03$, to describe this process above the convective regions. 

Finally, we activate the \verb|co_burn| net (parameter \verb|auto_extend_net = true|) after the main sequence to have a complete and extended network for C- and O-burning and $\alpha$-chains. Following \cite{simonato2025}, we define the core radius of each element as the radius inside which the fraction of the lighter element below $f_{\rm X}=10^{-4}$. For example, the helium (carbon) core radius is the radius inside which the mass fraction of H ($^{4}$He) drops below  $f_{\rm X}=10^{-4}$ ($f_{\rm Y}=10^{-4}$). Our definition is the same as the \textsc{MIST} libraries, and yields an accurate description of the core boundary \citep{Paxton2011,Paxton2013,Paxton2015,Dotter2016,Choi2016}.

\subsection{Initial conditions and stopping criteria}
We consider a grid of VMS models with masses from 50  to 500~M$_{\odot}$, by intervals of 25 M$_{\odot}$. We consider 14 metallicities from $Z = 10^{-4}$ to $Z=0.02$. The initial He mass fraction is $Y = Y_{\text{prim}} + Z \, \Delta Y / \Delta Z$, with $Y_{\text{prim}} = 0.24$ the primordial He abundance and $\Delta Y / \Delta Z$ is parametrized so that we range from a primordial ($Y = 0.24$, $Z = 0$) to nearly a solar abundance ($Y = 0.28$, $Z = 0.02$, \citealp{Pols1998})\footnote{Here, we refer to the traditionally considered value for solar metallicity $Z=0.02$ (\citealp{grevesse1998}; but see, e.g., \citealp{magg2022} for more recent estimates).}. Finally, we calculate the H mass fraction as $X = 1 - Y - Z$. We consider only non-rotating stellar models. 

We evolve the \textsc{mesa} models until the star exceeds a central temperature $\log{T_\mathrm{c}/\mathrm{K}}=9.55$, corresponding to the very last stages before core collapse, or it undergoes (pulsational) pair-instability. In the latter case, the creation of electron-positron pairs removes thermal pressure from the gas, possibly triggering a state of global dynamical instability. Because we use the hydro-static integrator of \textsc{mesa}, we cannot integrate the models until (pulsational) pair instability kicks in, but we verify a posteriori the global stability of the star. Specifically, we calculate the first adiabatic exponent $\Gamma_1$ (e.g., see also \citealp{Marchant2019,farmer19,Farmer2020,Costa2021}), properly weighted and integrated over the mass domain \citep{stothers1999}:
\begin{equation}
    \langle \Gamma_1 \rangle = \frac{\int_0^M\frac{P}{\rho} \, \Gamma_1 dm}{\int_0^M\frac{P}{\rho} \, dm}.
\end{equation}
Here, $P$ is the pressure, $\rho$ is the density and the integral is calculate up to the surface of the star. If a star enters a regime $\langle \Gamma_1 \rangle < 4/3 + 0.01$, we label it as pair-instability and no longer integrate its evolution.

\begin{figure*}
    \centering
    \includegraphics[width=\hsize]{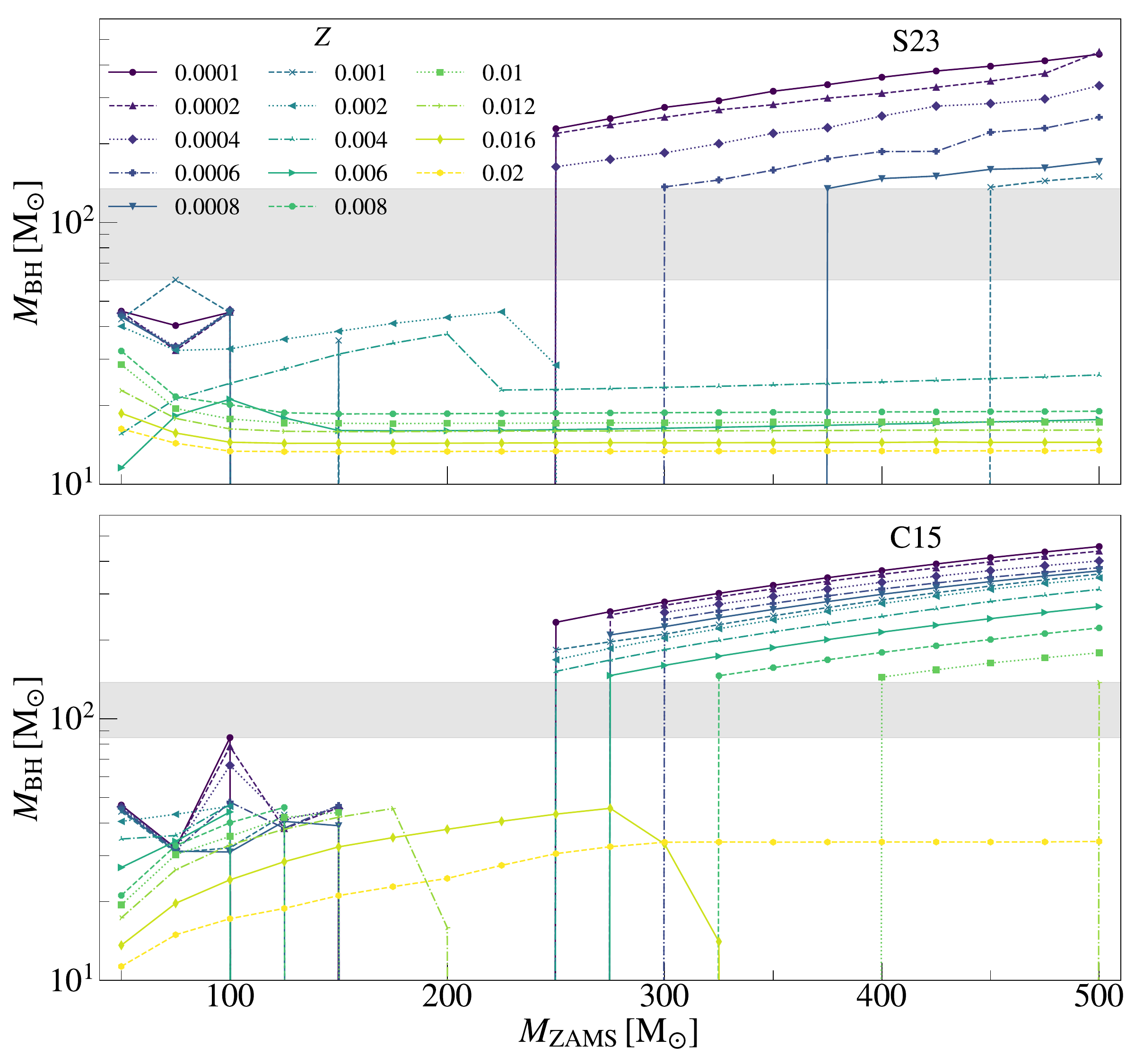}
    \caption{Final BH mass as a function of $M_{\mathrm{ZAMS}}$ for different metallicities $Z$,  for the \citetalias{Sabhahit2023} (upper) and \citetalias{Chen2015} (lower)  wind mass-loss prescription. The grey shaded area shows the resulting pair-instability mass gap for the two wind models.}
    \label{fig:MBH_Z}
\end{figure*}

\subsection{BH masses} \label{sec:bhmassmethods}
If the star enters the (pulsational) pair-instability regime, we calculate the final BH mass with the fitting formulas from 
\cite{Mapelli2020}, based on the models by \citet{Woosley2017}. 
In this model, a star undergoes a pulsational pair-instability supernova  
if the He core has a final mass in the range $32 \text{ M}_{\odot} \leq M_{\mathrm{He,f}} < 64 \text{ M}_{\odot}$. 
The mass of the BH is therefore estimated  as
\begin{equation}
M_{\mathrm{BH}} = \alpha_{\mathrm{P}}\,M_{\mathrm{CCSN}},
\end{equation}
where $M_{\mathrm{CCSN}}$ is the expected BH mass after a core-collapse supernova, based on the fitting formulas by \cite{Fryer2012}, and  $\alpha_{\mathrm{P}}$ depends on the mass of He-core and on the ratio between the He-core mass and the total mass in the pre-supernova stage. 

If $64 \text{ M}_{\odot} \leq M_{\mathrm{He,f}} \leq 135 \text{ M}_{\odot}$, instead, the star undergoes a pair-instability supernova,  
leaving no compact remnant. If the core mass exceeds $135 \, \msun$, pair instability triggers photodisintegration and the star directly collapses into a BH, leaving a compact remnant with $M_{\mathrm{BH}}$ equal to the total mass of the star.

\begin{figure}
    \centering
    \includegraphics[width=\hsize]{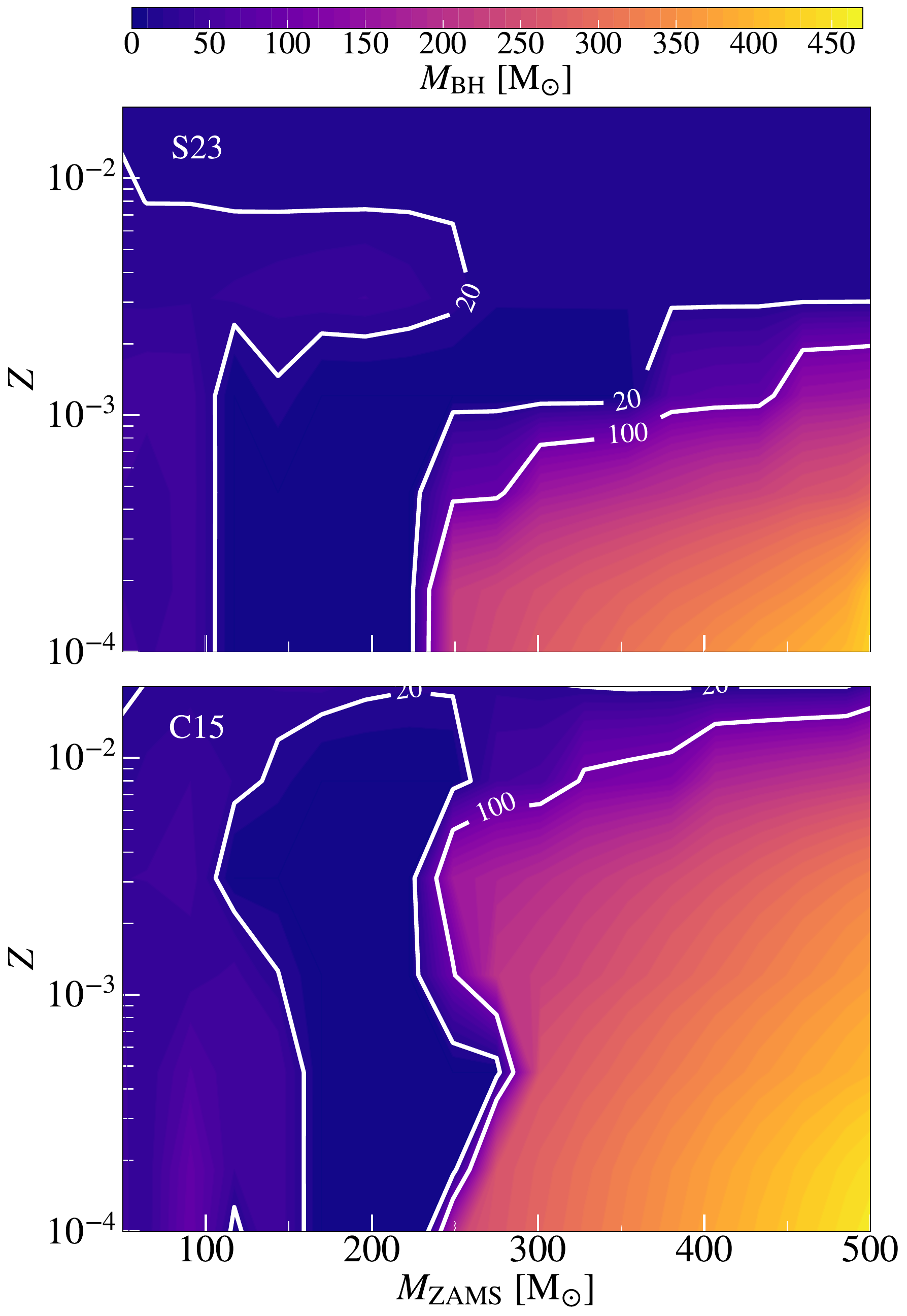}
    \caption{Contour plot showing the final BH mass as a function of $M_{\mathrm{ZAMS}}$ and $Z$, for the \citetalias{Sabhahit2023} (upper panel)  and the \citetalias{Chen2015} (lower panel) models. The white contours highlight the levels at $20 \, \msun$ and $100 \, \msun$.  
    }
    \label{fig:contour_BH}
\end{figure}

Mass ejection can also occur in stars that undergo a failed supernova because of the shock triggered by  instantaneous neutrino loss \citep{lovegrove2013,fernandez2018}. The ejected mass depends on the ejected energy and the compactness of the star \citep{fernandez2018}. In particular, the ejected energy depends on the core compactness, $\xi_{\mathrm{2.5}}$ \citep{oconnor2011, burrows2021,burrows2024}.
Also, the ejected mass decreases with the stellar envelope compactness, $\xi_{\mathrm{env}} = M_* / R_*$,
where $R_*$ is the total radius of the star. The ejected mass depends on $\xi_{\mathrm{env}}$ \citep{fernandez2018}, and ranges from $M_{\mathrm{ej}}\sim 10 \, \msun$ for red supergiant stars ($\xi_{\mathrm{env}}\sim 10^{-2}$)  to $M_{\mathrm{ej}}\lesssim 10^{-3} \, \msun$ if the star dies as a Wolf-Rayet object ($\xi_{\mathrm{env}}\sim 10$).

Following \cite{Costa2022}, we estimate the BH mass for the stars that do not undergo a (pulsational) pair instability supernova as:
\begin{equation}
    M_{\mathrm{BH}} = M_* - \delta M_{\mathrm{G}} - M_{\mathrm{ej}},
\end{equation}
where $M_*$ is the final stellar mass,  $\delta M_{\mathrm{G}}$  is the instantaneous gravitational mass loss, and $M_{\mathrm{ej}}$ is the the mass induced by neutrino-driven shocks in a failed supernova. Here, we assume $\delta M_{\mathrm{G}} = 0.3 \, M_{\odot}$ \citep{fernandez2018,costa2025}, which typically yields a corresponding ejected energy between $10^{47}$ and $5 \times 10^{47}$ erg. We thus calculate $M_{\mathrm{ej}}$ as the mass of the star with a binding energy less than $3 \times 10^{47}$ erg. 

\subsection{Population-synthesis with \textsc{sevn}}\label{sec:sevn}
We explore the effect of optically thick winds on the mass spectrum of BHs, BBHs and BBH mergers using the population-synthesis code \textsc{sevn} (\citealp{Spera2017, Spera2019, Mapelli2020,Iorio2023})\footnote{
We used  version 2.10.1 of \textsc{sevn}, updated to commit \href{https://gitlab.com/sevncodes/sevn/-/commit/c9dc8e4578990fd85cba91e082ebff6068388f56}{c9dc8e457}. \textsc{sevn} is publicly available  \href{https://gitlab.com/sevncodes/sevn}{{here}}.}. \textsc{sevn} interpolates pre-computed stellar tracks to evolve single and binary stars. Also, it incorporates semi-analytic formulas for processes like binary mass transfer, tides, supernova explosion, and gravitational-wave decay. We refer to \citet{Iorio2023} for a complete description of the code.

We generate new evolutionary tables for stars with mass $M_{\star} \geq 50\,{} \mathrm{ M}_{\odot}$ from the aforementioned VMS models with \textsc{mesa}. We then evolve single and binary stars with primary mass in this mass range. 
For binary stars, we use the stellar tables from \textsc{parsec} \citep{Bressan2012,Chen2015, Costa2019b, Costa2019a, Nguyen2022,costa2025} to account for the evolution of secondary stars with mass $M_{\star} < 50 \, \msun$.

When two stars collide, we use the standard formalism of \textsc{sevn} as described by \cite{Iorio2023}. Specifically, we assume that the collision product retains the total mass of the two stars. The CO and He core masses of the collision product are the sum of the CO and He core masses of the two colliding stars. Finally, the collision product inherits the phase and percentage of life of the most evolved progenitor star. Hence, our models likely overestimate the total mass of the collision product and underestimate mixing. For a more self-consistent treatment of massive star collisions, see, e.g., \cite{costa2022a} and \cite{Ballone2023}.

We  randomly draw the zero-age main sequence (ZAMS)  primary star masses from a \cite{Kroupa2001} initial mass function, with $M_\mathrm{ZAMS} \in \ [50,500]$. For binaries, we draw the secondary masses from the observation-based mass ratio distribution  $\mathcal{F}(q) \propto q^{-0.1}, \;  \mathrm{with} \; q \; \in [0.1,\,{}1]$ from  \cite{sana2012}. We generate the initial orbital periods ($P$) and eccentricities ($e$) from the distributions by \cite{sana2012}: 
$\mathcal{F}(\mathcal{P}) \propto \mathcal{P}^{-0.55}, \; \mathrm{with} \; \mathcal{P}=\log_{10}(P/ {\rm days})\in [0.15,\,{}5.5]$ and $\mathcal{F}(e) \propto e^{-0.45}, \; \mathrm{with} \; e \in [10^{-5},\,{}e_{\rm max} (P)]$, respectively.
For a given orbital period, we set the upper limit of the eccentricity distribution $e_{\rm max} (P)$ as $e_{\rm max} (P) = 1 - \left[ P / (2 \, {\rm days}) \right]^{-2/3}$ \citep{moe2017}.

We investigate eight different metallicities: $Z = 0.0001$, 0.001, 0.002, 0.004, 0.006, 0.008, 0.01, 0.02. For each metallicity, we generate $2\times10^6$ single and $1\times10^7$ binary systems. Our initial setup corresponds to the same as the fiducial model by \cite{Iorio2023}. 

{To highlight the impact of stellar winds on the resulting BH mass distribution, we also perform a run with tables from the non-rotating \textsc{parsec} tracks presented by \cite{costa2025}\footnote{The \textsc{parsec} stellar tracks are publicly available at \url{http://stev.oapd.inaf.it/PARSEC/} \citep{costa2025}.}. In this case, 
the adopted wind model is the same as described by \citet[][hereafter  \citetalias{Chen2015}]{Chen2015}.
Specifically, the winds of massive O-type stars are modeled adopting the fitting formulas by \citet{Vink2000} and \cite{Vink2001}, with the correction by \citet{Graefener2008} to account for the effects of electron scattering. The mass-loss rate of an O-type star scales with the metallicity as $\dot{M} \propto Z^{\beta}$, where $\beta$ depends on the Eddington ratio $\Gamma_{\rm Edd}$:
$\beta = 0.85$ for $\Gamma_{\rm Edd} < 2/3$,
$\beta = 2.45 - 2.4\times\Gamma_{\rm Edd}$ for $2/3 \le \Gamma_{\rm Edd} < 1$,
and $\beta = 0.05$ for $\Gamma_{\rm Edd} \ge 1$ \citep{Chen2015}.
In this way, this formalism incorporates the dependence of mass loss on the Eddington ratio \citep{Graefener2008,Vink2011}. As for Wolf–Rayet (WR) stars, \textsc{parsec} adopts the prescriptions by \citet{Sander2019}, which 
match the  properties of Galactic WR stars of type C (WC) and type O (WO). We refer to \cite{costa2025} for other details on these stellar tracks.}

\section{Results} \label{sec:results}

\subsection{Stellar cores}
Figure \ref{fig:pisn_new} shows the final He and CO core masses for the \citetalias{Sabhahit2023} and \citetalias{Chen2015} wind models. At  low metallicity ($Z \leq{} 0.0001$), the two wind models predict the same monotonic increase of the He and CO core mass with $M_{\mathrm{ZAMS}}$, because VMSs do not enter the optically thick wind regime. At these metallicities, the \citetalias{Chen2015} models generally yield slightly smaller cores, due to the different envelope undershooting and convection models \citep{simonato2025}. 

At higher metallicity, the two considered models \citetalias{Chen2015} and \citetalias{Sabhahit2023}  behave in a dramatically different way. According to  \citetalias{Sabhahit2023}, VMSs can enter the optically thick wind regime at $Z\ge{}0.001$: such strong winds reduce the final total mass and core mass of VMSs, resulting in a flat or even decreasing trend of $M_{\rm c}$ with $M_{\rm ZAMS}$. Hence, the \citetalias{Sabhahit2023} model predicts smaller He and CO cores compared to \citetalias{Chen2015}, with a deep impact on the occurrence of (pulsational) pair instability. 

The effect of  optically thick winds in the \citetalias{Sabhahit2023} model is twofold. On the one hand, the \citetalias{Sabhahit2023} model results in a lower metallicity threshold below which VMSs can enter the (pulsational) pair instability regime: the threshold is $Z_{\rm th}=0.004$ and $0.01$ for \citetalias{Sabhahit2023} and \citetalias{Chen2015}, respectively. On the other hand, the formation of IMBHs via direct collapse is suppressed in the \citetalias{Sabhahit2023} model for metallicity $Z\geq{}0.001$, whereas the \citetalias{Chen2015} model allows the formation of IMBHs already at $Z\leq{}0.008$ because of the lower wind mass-loss.

For example, at $Z = 0.001$ optically thick winds become effective for $M_{\mathrm{ZAMS}}> 200 \, \msun $, when the stellar core is already well within the pair-instability regime. Mass loss prevents the He core mass from increasing above $135 \, \msun$. 
As a consequence, only VMSs with $M_{\mathrm{ZAMS}} \gtrsim  450 \, \msun $ undergo direct collapse to IMBH in the \citetalias{Sabhahit2023} model, whereas in \citetalias{Chen2015} IMBHs can form via direct collapse already at $M_{\mathrm{ZAMS}} \sim 300 \, \msun $. 

For $Z=0.004$, VMSs with $M_{\mathrm{ZAMS}} \geq 200 \,{}\msun$ develop stellar winds that are strong enough to reduce the stellar core to $M_{\mathrm{c}} < 30 \,{}\msun$, where even pulsational pair instability is inhibited.

For $Z \gtrsim 0.006$, VMSs in \citetalias{Sabhahit2023} models undergo optically thick stellar winds throughout the mass range considered. This dramatically reduces final core masses, which display a weak dependence on $M_{\mathrm{ZAMS}}$. For example, at $Z \sim 0.006$ the core mass for VMSs with $M_{\mathrm{ZAMS}} > 400 \, \msun$ decreases from $\sim 200 \, \msun$ in the \citetalias{Chen2015} model to $\sim 20 \, \msun$ in \citetalias{Sabhahit2023}.

\subsection{BH masses}
Figure \ref{fig:MBH_Z} shows the final BH mass as a function of $M_{\mathrm{ZAMS}}$ for models \citetalias{Chen2015} and \citetalias{Sabhahit2023}. The key difference between the two models is the metallicity threshold to form IMBHs; this is $Z_{\rm IMBH}=0.001$ and 0.012 for models \citetalias{Sabhahit2023} and  \citetalias{Chen2015},  respectively. 
In model \citetalias{Sabhahit2023}, IMBH formation is suppressed at metallicity $Z>{0.001}$ by the onset of optically thick winds. These strong winds erode the mass of the He core and prevent it from reaching the threshold $M_{\rm He}\sim{135}$~M$_\odot$, above which the star avoids to explode as a  pair-instability supernova. Thus, even the most massive stars do not produce IMBHs in the \citetalias{Sabhahit2023} model unless they have metallicity $Z\leq{}0.001$. In contrast, the lower mass loss in model \citetalias{Chen2015} allows to grow larger cores and enables the formation of IMBHs up to $Z\sim{0.012}$.

In absence of optically thick winds, the BH mass increases with $M_{\mathrm{ZAMS}}$, up to the onset of (pulsational) pair instability, at $M_{\mathrm{ZAMS}} \gtrsim 75 \, \msun $. For larger initial masses, the onset of direct collapse depends on metallicity. 
In the \citetalias{Chen2015} model, VMSs undergo (pulsational) pair instability up to $Z=0.01$. At lower metallicities, all the models with $M_{\mathrm{ZAMS}}> 400 \, \msun$ produce IMBHs, with masses ranging from $135 \, \msun$ to $450 \, \msun$. At $Z=0.02$, instead, wind mass-loss is strong enough to prevent the onset of pair instability, leading to a maximum BH mass $M_{\mathrm{BH}}\sim 30 \, \msun$. 

In the \citetalias{Sabhahit2023} model, the maximum metallicity at which IMBHs form is an order of magnitude lower ($Z=0.001$) compared to \citetalias{Chen2015}. Below this threshold, the \citetalias{Sabhahit2023} model produces IMBHs with masses in the same  range as \citetalias{Chen2015}. 
At higher metallicities, the most massive BHs originate from stars with $M_{\mathrm{ZAMS}} \sim{50} \msun$, while more massive stars  deliver $M_{\mathrm{BH}} \lesssim 15 \msun$. This happens because VMSs develop optically thick winds at an early stage, and spend all their lifetime with $\dot M>10^{-5}$ M$_\odot$ yr$^{-1}$. This results in a severe erosion of the core \citep{simonato2025,boco2025}.  Overall, the \citetalias{Sabhahit2023} models predict a pair-instability BH mass gap from $65 \, \msun$ to  $135 \, \msun$.

Figure \ref{fig:contour_BH} shows the final BH mass as a function of the ZAMS mass and the metallicity for the \citetalias{Sabhahit2023} model. 
The formation of IMBHs takes place only for VMSs with $M_{\mathrm{ZAMS}} \gtrsim 250 \, \msun$ at $Z \leq 0.001$. 
Also, the formation of BHs with $M_{\mathrm{BH}} > 20\,{}\msun$ is suppressed for $Z>0.01$ throughout the mass range.

\begin{figure}
    \centering
    \includegraphics[width=\hsize]{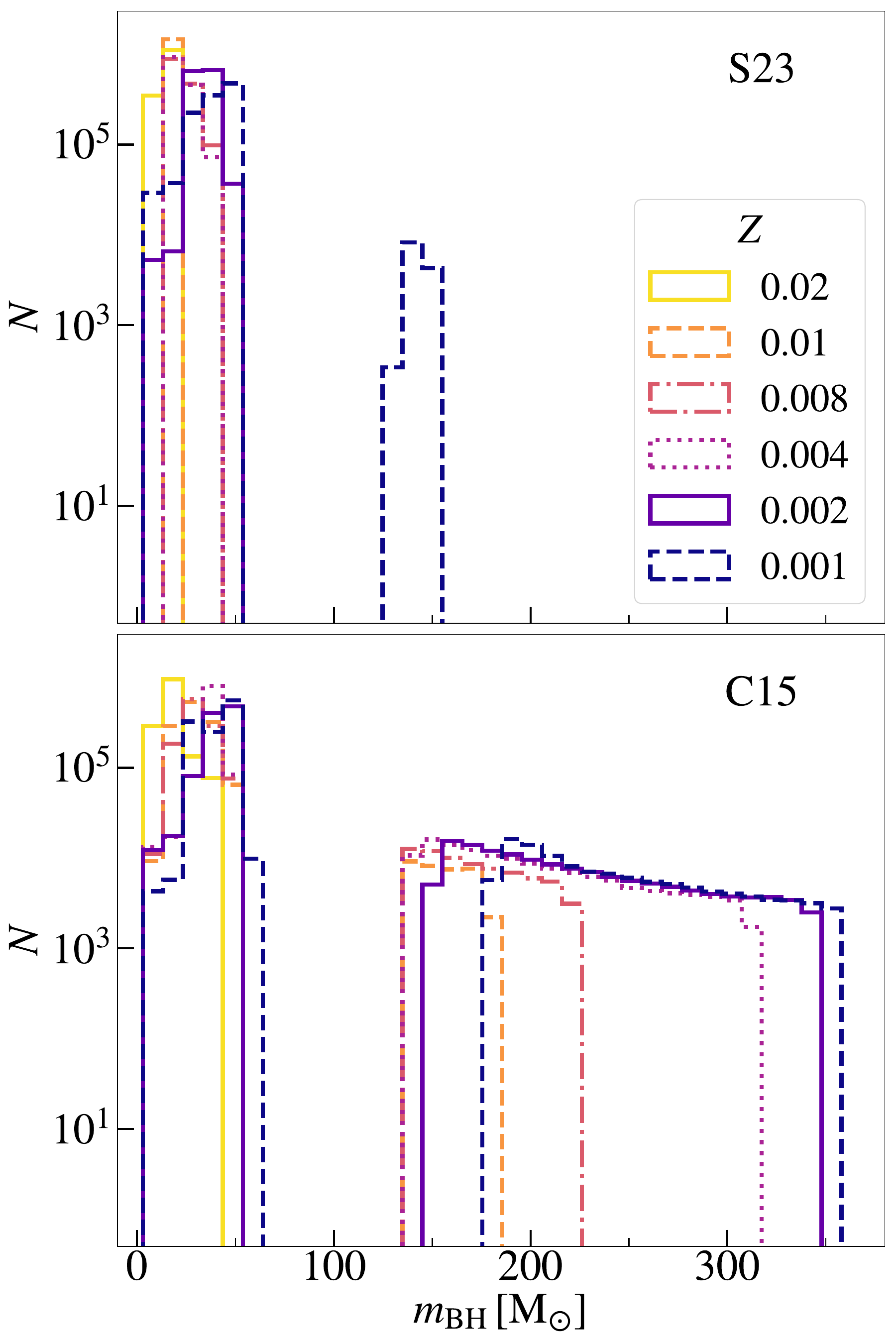}
    \caption{BH mass distribution from single stars with $M_{\mathrm{ZAMS}} > 50 \,{}\msun$, for the models adopting the \citetalias{Sabhahit2023} (upper panel) and \citetalias{Chen2015} (lower panel) wind prescriptions and for different metallicities. 
    }
    \label{fig:single_BH_pop}
\end{figure}

\begin{figure*}
    \centering
    \includegraphics[width=\textwidth]{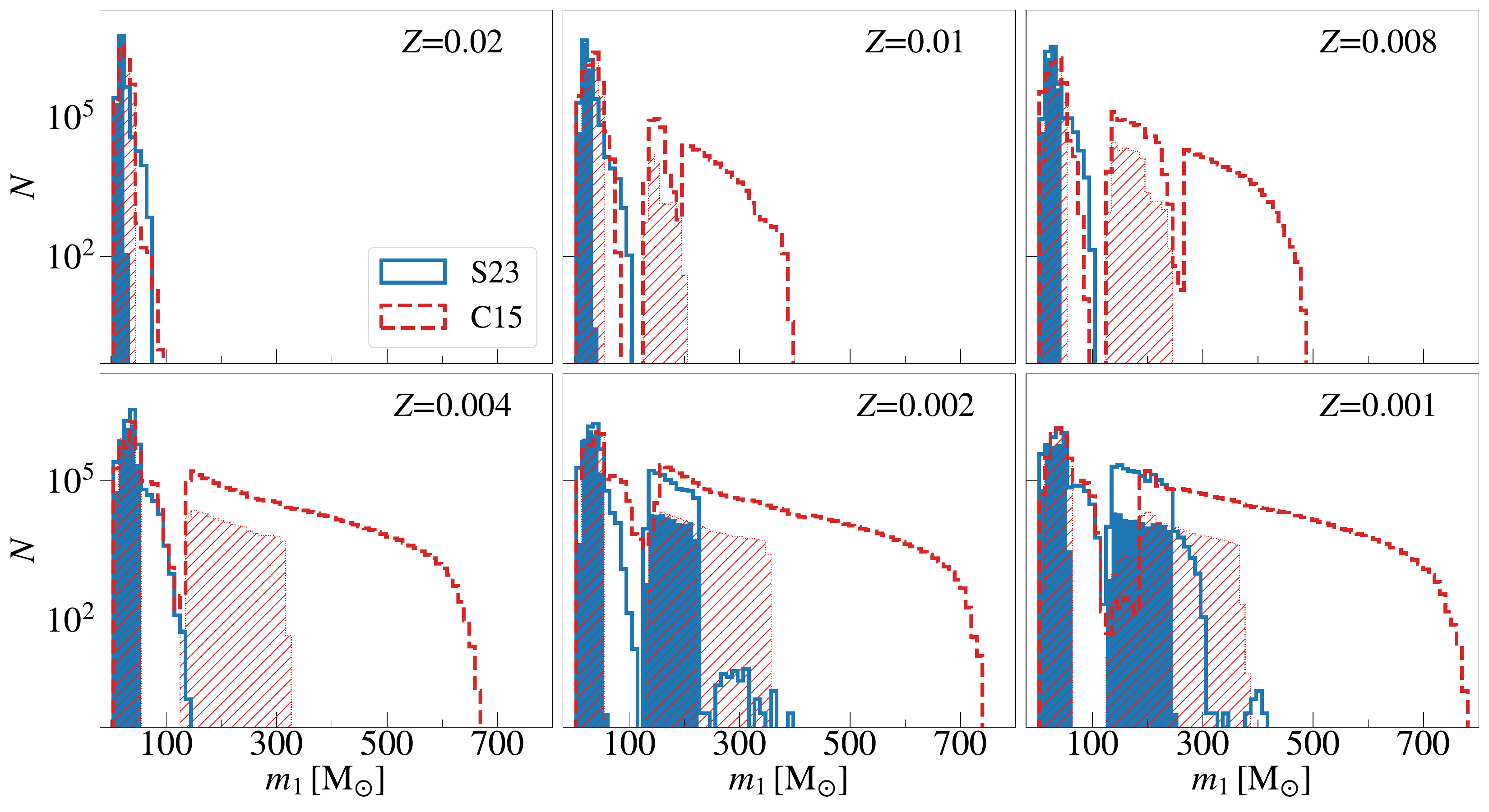}
    \caption{BH mass distribution from binaries with $M_{\mathrm{ZAMS,1}} > 50 \msun$, for models adopting the \citetalias{Sabhahit2023} (blue, solid line, solid fill) and \citetalias{Chen2015} (orange, dashed line, hatched area) wind prescriptions. The shaded (hatched) areas indicate the BHs in BBHs, while the empty histograms display the single BHs from disrupted binaries, including BHs that form from stellar collisions.} 
    \label{fig:bin_BH_pop}
\end{figure*}

\subsection{BH mass function}

\subsubsection{BHs from single VMSs}
Figure \ref{fig:single_BH_pop} shows the BH mass function from single VMSs assuming a Kroupa \citep{Kroupa2001} initial mass function. For $Z > 0.001$  optically thick winds strongly suppress the production of IMBHs from single stars. At $Z=0.001$, the \citetalias{Sabhahit2023} model still allows  IMBH formation, but the efficient mass loss limits the maximum BH mass to $150 \, \msun$, compared to $350 \, \msun$ in the \citetalias{Chen2015} model.
For higher metallicities, stellar winds progressively reduce the maximum BH mass in the \citetalias{Sabhahit2023} model, down to $20 \, \msun$ at $Z=0.02$.

In the \citetalias{Chen2015} model, the formation of IMBHs takes place up to $Z = 0.01$, where the maximum BH mass is $\sim 180 \, \msun$ (see also \citealp{costa2025}). At lower metallicities, the maximum IMBH mass reaches $\sim{400}$ M$_\odot$, for ZAMS masses up to M$_{\rm ZAMS}=500$~M$_\odot$. In this model, IMBH formation from single VMSs is only suppressed at $Z = 0.02$, where BHs can form with a mass up to $33 \, \msun$.

\subsubsection{BBHs from  VMS binaries}
Figure \ref{fig:bin_BH_pop} shows the primary BH masses from binaries with primary mass $M_{\mathrm{ZAMS,1}} > 50 \, \msun$, distributed according to a Kroupa \citep{Kroupa2001} mass function. We refer to Section~\ref{sec:sevn} for more detail on the initial conditions.  We display the mass distribution of both primary BHs in BBHs and single BHs from disrupted binaries. Star-star collisions dramatically extend the BH mass function to larger values compared to the single stellar-evolution case in Fig. \ref{fig:single_BH_pop}.

The \citetalias{Chen2015} model predicts the formation of BHs as massive as $\gtrsim 700 \, \msun$ for $Z \leq 0.002$. Also, IMBHs up to $\sim 400 \, \msun$ can still form at $Z = 0.01$. 
In the \citetalias{Sabhahit2023} model, instead, the formation of IMBHs is suppressed already at $Z = 0.004$, where they represent less than 0.2\% of the total BH number, despite the occurrence of star-star colllisions. In contrast, the percentage of IMBHs is $21\%$  in the \citetalias{Chen2015} model at $Z = 0.004$. We find that, in the metallicity range $Z=0.004-0.008$, our distributions are consistent with those from \cite{shepherd2025} when they consider the same wind prescription.

Primary BHs in BBHs also reach higher masses than BHs from single stellar evolution, due to mass accretion from the stellar companion. 
In the \citetalias{Sabhahit2023} model, IMBH formation in BBHs is possible for $Z\leq0.002$, while for $Z=0.004$, the maximum primary BH mass in BBHs  decreases to $45 \, \msun$. At $Z=0.02$, the BH mass function in the \citetalias{Sabhahit2023} model extends from $5 \, \msun$ to $20 \, \msun$. 

BHs in BBHs display a pair-instability mass gap between $65 \, \msun$ and $135 \, \msun$ for both the \citetalias{Chen2015} and the \citetalias{Sabhahit2023} models. In contrast, star-star collisions can produce BHs in this mass range, and fill the pair-instability mass gap with single BHs.

\subsubsection{BBH mergers}
\begin{figure}
    \centering
    \includegraphics[width=\hsize]{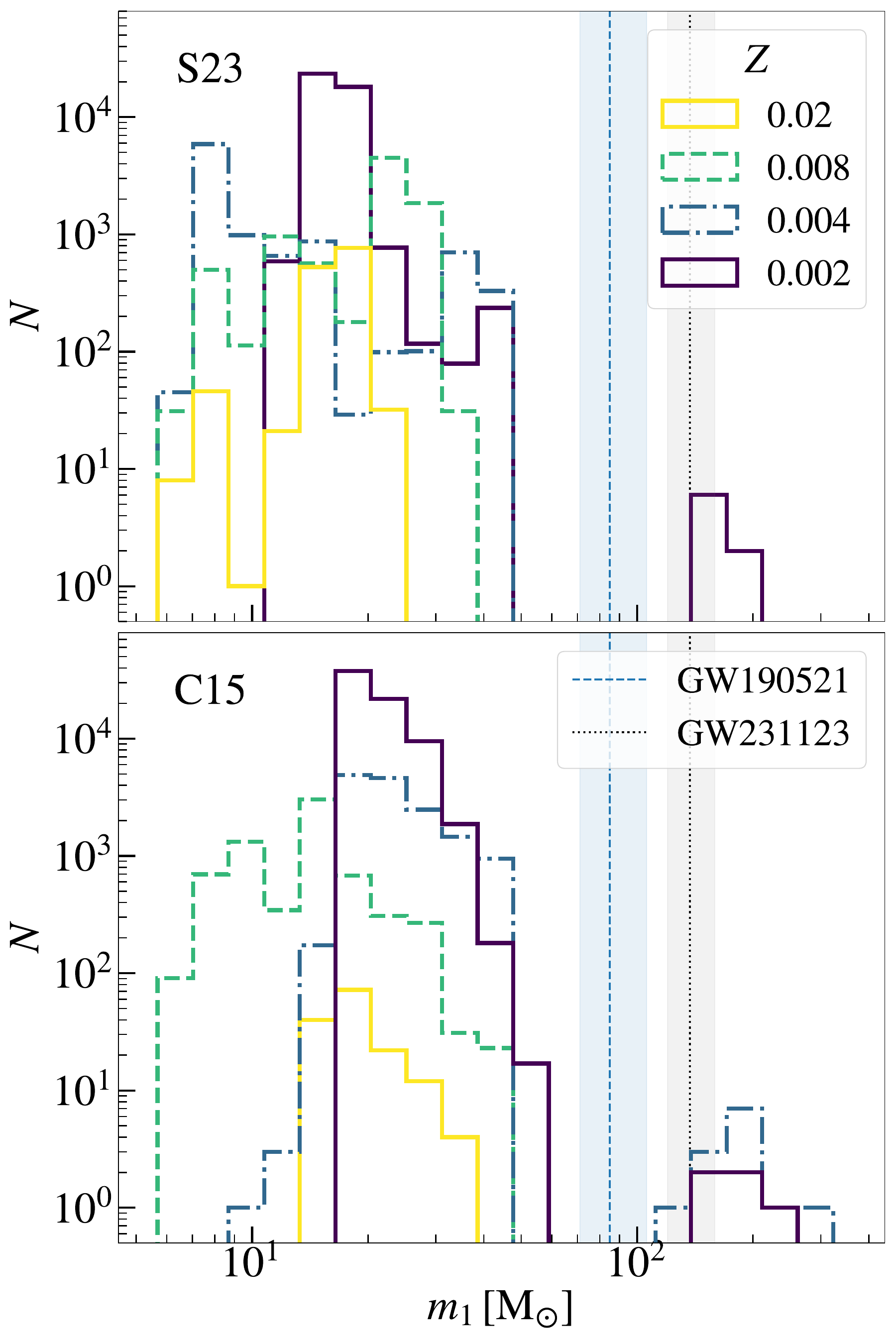}
    \caption{Primary BH mass distribution in BBH mergers with $M_{\mathrm{ZAMS,1}} > 50 \msun$, for the \citetalias{Sabhahit2023} (upper) and \citetalias{Chen2015} (lower) wind prescriptions, and for different metallicities. The blue dashed (black dotted) line displays the inferred value for GW190521 (GW231123), with the estimated uncertainty (shaded area). 
    }
    \label{fig:mergers_BH_pop}
\end{figure}

\begin{figure}
    \centering
    \includegraphics[width=\hsize]{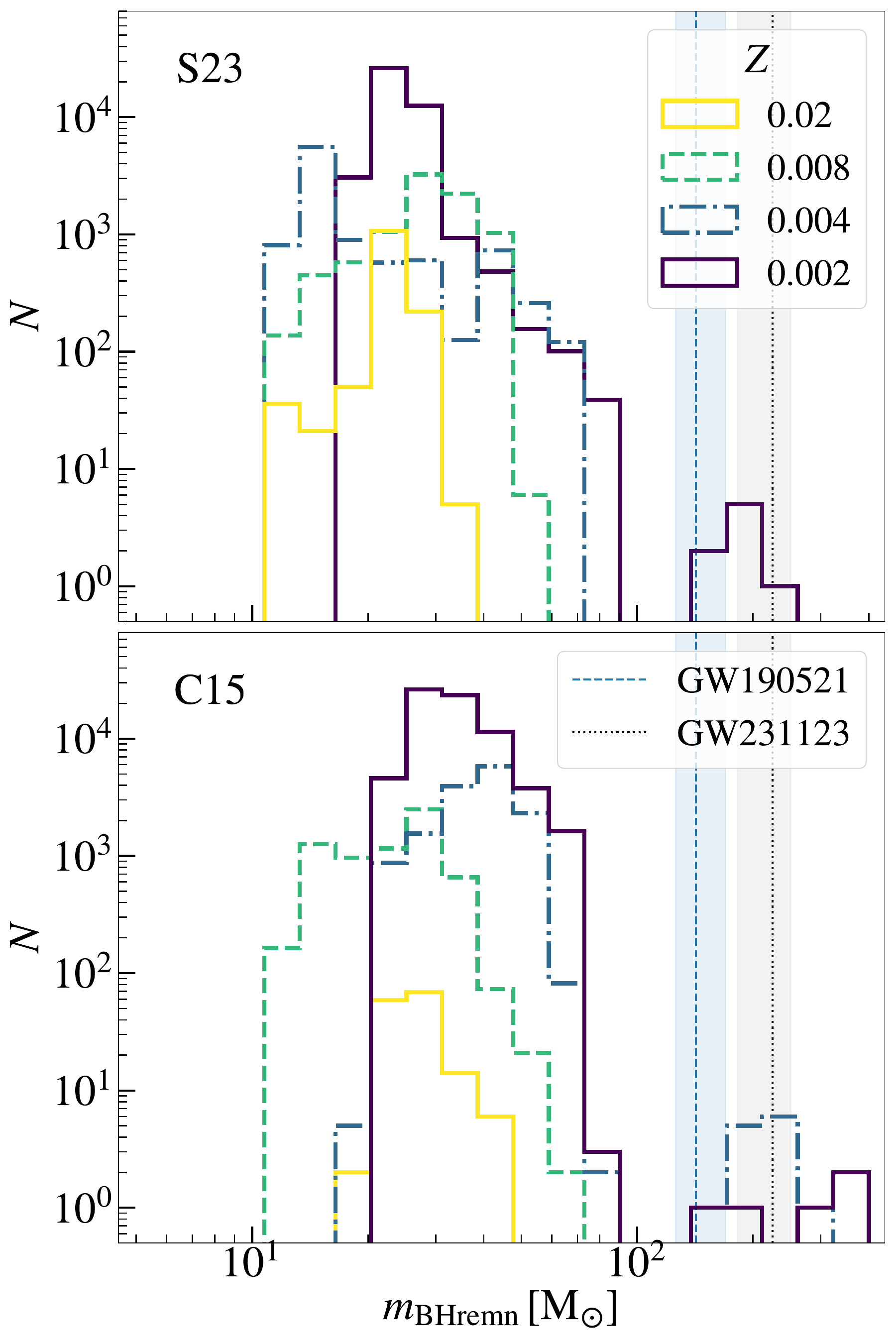}
    \caption{Same as Fig.~\ref{fig:mergers_BH_pop}, but for the resulting BH remnant mass.
    }
    \label{fig:mergers_mtot_BH_pop}
\end{figure}

Figures \ref{fig:mergers_BH_pop} and  \ref{fig:mergers_mtot_BH_pop} show the primary and remnant mass distributions for BBHs that merge within a Hubble time. In the \citetalias{Sabhahit2023} model, no BBH merger has a primary mass $m_1 > 50 \, \msun$ for $Z>0.002$. In the \citetalias{Chen2015} case, IMBH mergers can occur up to a metallicity of  $Z=0.004$. In general, the BBH merger mass distribution displays a mild dependence on metallicity, with a peak at $\sim 10 \, \msun$. 

The maximum merger remnant produced by these BBHs is usually two times as massive as the maximum primary BH mass. In presence of optically thick winds (model \citetalias{Sabhahit2023}), the formation of IMBHs from BBH mergers is thus possible up to $Z=0.002$, where $m_{\mathrm{BH,max}} \sim 240 \, \msun$. At higher metallicities, the maximum BH remnant mass progressively decreases down to $35 \, \msun$ at $Z=0.02$. 

In the \citetalias{Chen2015} models, the masses delivered by BBH mergers are larger, with the possibility of having IMBH remnants up to a metallicity of $Z=0.006$. 

\subsubsection{Progenitors of GW231123 and GW190521}

We qualitatively compare our models with the most massive BBH mergers detected by the LIGO-Virgo-KAGRA interferometers, namely GW231123 \citep{gw231123} and GW190521 \citep{abbottGW190521,abbottGW190521astro}. 
According to our \citetalias{Sabhahit2023} model, the progenitors of GW231123 should have had a metallicity $Z \leq 0.002$. 
The primary mass of GW190521 \citep{abbottGW190521,abbottGW190521astro}, instead, lies well within the mass gap predicted by both our \citetalias{Chen2015} and \citetalias{Sabhahit2023} models. In a follow-up work, we will explore the dynamical formation of such mergers with our new \citetalias{Sabhahit2023} models (Torniamenti et al., in prep.).

\section{Discussion} \label{sec:discussion}

\subsection{Impact of rotation and overshooting}

{In this work, we investigate the impact of a recent  model accounting for optically thick winds \citepalias{Sabhahit2023} on the formation of IMBHs from VMSs. We have considered non-rotating stellar models with a fixed overshooting parameter. However, both stellar rotation and core overshooting may have a relevant impact on the final BH mass. In a follow-up paper, we will include these additional parameters in detail. In the following, we briefly discuss their expected impact on our results.}

{Stellar rotation affects both the envelope and the core mass, leading to enhanced stellar-mass loss during the main sequence (\citealp{langer1998}, but see also \citealp{muller2014}), and triggering chemical mixing in the core \citep{Sabhahit2023}. In the optically thin regime, rotation generally results in enhanced mass loss when the critical rotation is approached ($\Omega/\Omega_{\mathrm{c}}>0.4$). As a consequence, it can further quench the BH production at low metallicity $Z<0.01 \, Z_{\mathrm{\odot}}$, where optically thick winds are not activated  \citep{winch2024,winch2025}. Also, rotation can drive the transition to the optically-thick regime at lower initial masses \citep{Sabhahit2023,boco2025}. For VMS, this mainly affects the evolution at low $Z$, where  the star can retain a large fraction of its angular momentum due to weaker mass loss. At high $Z$, wind mass loss can efficiently remove the angular momentum, leading to a minor impact from stellar rotation on the final BH mass.}

{Core overshooting also triggers chemical mixing, affecting the final core mass similarly to rotation \citep{Sabhahit2022}. In the optically-thin regime, increasing the overshooting parameter leads to more massive cores, with larger values $M_{\mathrm{BH}}$ for the same $M_{\mathrm{ZAMS}}$ \citep{winch2024}. For VMSs, optically thick winds already lead to the formation of a fully mixed star if the initial stellar mass is large enough ($M_{\mathrm{ZAMS}}>200 \, \msun$, \citealp{Sabhahit2022}), while removing the stellar envelope. As a consequence, overshooting is not expected to play an important  role on the final BH mass. }

{In summary, we expect overshooting and rotation to affect the relation between $M_{\mathrm{ZAMS}}$ and $M_{\mathrm{BH}}$ in the low-metallicity and low-mass end of the parameter space considered, where mass loss is driven by optically-thin winds. In such regimes, a larger overshooting parameter and a fast rotation 
may further reduce the maximum value of $Z$ at which IMBHs are produced. When more {vigorous} stellar winds are present, VMS evolution is expected to mainly be  driven by stellar mass loss, with a less relevant impact from these parameters. 
}

\subsection{Caveats of VMS population models}
{In our population-synthesis models, we have simply drawn primary stellar masses from a Kroupa IMF between 50 and 500 M$_\odot$, corresponding to the extreme high-mass end of the mass function. 
In most star forming regions, the presence of stars in this mass range is dominated by stochastic effects or possibly by additional quenching effects \citep[e.g.,][]{Weidner2006}. Thus, the conclusions of our population-synthesis simulations only apply to extreme star-forming regions, such as the R136  cluster in the LMC  \citep{Crowther2010,bestenlehner2020}, the super-star clusters in starburst galaxies \citep[e.g., the Antennae galaxies,][]{Whitmore1995} and the high-redshift massive clusters recently discovered by the James Webb Space Telescope \citep{Vanzella2023,Adamo2024, Mowla2024}. Moreover, it is also possible that VMSs are the result of  multiple stellar collisions in very dense star clusters \citep{Portegies2002}. In such case, population studies require dedicated $N-$body simulations \citep[e.g.,][]{dicarlo2021, Rantala2024}.}

\section{Conclusions} \label{sec:summary}

We have investigated the impact of a new optically thick wind model \citepalias{Sabhahit2023} on the formation of IMBHs in {non-rotating} isolated single and binary very massive stars (VMSs).  The new model is supported by several  observational constraints, such as the pair-instability supernova rate \citep{simonato2025} and the existence of single Wolf-Rayet stars at low metallicity \citep{boco2025}. 
We have integrated the evolution of VMSs with \textsc{mesa}  \citep{Paxton2019} and have folded our new \textsc{mesa} tracks inside our binary population synthesis code \textsc{sevn}  \citep{Iorio2023}.

We find that IMBH production is suppressed at intermediate metallicity ($Z\in{}[10^{-3},0.01]$), compared to less {effective}  wind models. 
According to our models, IMBHs cannot form from VMS collapse at the metallicity of the LMC ($Z\sim{0.008}$) and possibly even the SMC  ($Z\sim{0.0025}$).  

IMBHs might be able to form   via stellar collisions at metallicity as high as $Z = 0.004$, but only under the very optimistic assumption that each collision triggers no mass loss.
However, even with this assumption for stellar collisions, the occurrence fraction of IMBHs over the total number of BHs ($\lesssim0.2\%$) is two orders of magnitude lower  compared to models with less {effective wind mass-loss}. 

BBH mergers can produce IMBH remnants at metallicity $Z \leq 0.002$. At higher metallicities, the enhanced mass-loss produces BBH mergers with primary mass $m_{\mathrm{1}}< 50 \, \msun$, making it impossible to form IMBHs from first-generation BH mergers.

The BH mass spectrum from single VMSs shows a pair-instability mass gap from $65 \, \msun$ to  $135 \, \msun$. Such mass gap is still present in the mass distribution of BHs in BBHs and BBH mergers.

According to our \citetalias{Sabhahit2023} models, the progenitors of the most massive BBH observed by the LIGO-Virgo-KAGRA collaboration, GW231123 \citep{gw231123} must have had a  metallicity $Z \leq 0.002$. In a follow-up study, we will explore the impact of star cluster dynamics on this result.

\begin{acknowledgements}
{We thank the anonymous referee for their insightful comments and suggestions.} We thank Gautham Sabhahit and collaborators for sharing their \textsc{mesa} inlists (\url{https://github.com/Apophis-1/VMS_Paper1}, \url{https://github.com/Apophis-1/VMS_Paper2}). We thank Guglielmo Costa and Kendall Shepherd for useful comments and discussions.
ST acknowledges financial support from the Alexander von Humboldt Foundation for the Humboldt Research Fellowship.
ST, MM and GI acknowledge financial support from the European Research Council for the ERC Consolidator grant DEMOBLACK, under contract no. 770017. ST, MM and LB also acknowledge financial support from the German Excellence Strategy via the Heidelberg Cluster of Excellence (EXC 2181 - 390900948) STRUCTURES.  
GI  acknowledges financial support from the La Caixa Foundation for the La Caixa Junior Leader fellowship 2024.
GI also acknowledges financial support under the National Recovery and Resilience Plan (NRRP), Mission 4, Component 2, Investment 1.4, - Call for tender No. 3138 of 18/12/2021 of Italian Ministry of University and Research funded by the European Union – NextGenerationEU. EK and MM acknowledge support from the PRIN grant METE under contract No. 2020KB33TP. EK acknowledges financial support from the Fondazioni Gini for the Gini grant. The authors acknowledge support by the state of Baden-W\"urttemberg through bwHPC and the German Research Foundation (DFG) through grants INST 35/1597-1 FUGG and INST 35/1503-1 FUGG.
We use the \textsc{mesa} software (\url{https://docs.mesastar.org/en/latest/}) version r12115; \citep{Paxton2011, Paxton2013, Paxton2015, Paxton2018, Paxton2019}. We use \textsc{sevn} (\url{https://gitlab.com/sevncodes/sevn}) to generate our BBHs catalogs \citep{Spera2019,Mapelli2020,Iorio2023}, \textsc{trackcruncher} (\url{https://gitlab.com/sevncodes/trackcruncher}) \citep{Iorio2023} to produce the tables for the interpolation. The \textsc{parsec} stellar tracks that we used for this work are publicly available at \url{http://stev.oapd.inaf.it/PARSEC/} \citep{costa2025}. This research made use of \textsc{NumPy} \citep{Harris20}, \textsc{Pandas} \citep{Pandas2024}, \textsc{SciPy} \citep{SciPy2020} and \textsc{Matplotlib} \citep{Hunter2007}.

\end{acknowledgements}

\bibliographystyle{aa}
\bibliography{references}

\end{document}